# Intersubband polariton-polariton scattering in a dispersive microcavity

M. Knorr[1], J.M. Manceau[2,*], J. Mornhinweg[1], J. Nespolo[3], G. Biasiol[4], N.L. Tran[2], M. Malerba[2], P. Goulain[2], X. Lafosse[2], M. Jeannin[2], M. Stefinger[1], I. Carusotto[3], C. Lange[5,*], R. Colombelli[2], R. Huber[1]

[1]Department of Physics, University of Regensburg, 93053 Regensburg, Germany

[2]Centre de Nanosciences et de Nanotechnologies (C2N), CNRS UMR 9001, Université Paris Saclay, 91120 Palaiseau, France

[3]INO-CNR BEC Center and Dipartimento di Fisica, Universita di Trento, I-38123 Povo, Italy

[4]Laboratorio TASC, CNR-IOM, Area Science Park, 34149, Basovizza, Trieste, Italy

[5]Department of Physics, TU Dortmund University, 44227 Dortmund, Germany

**The ultrafast scattering dynamics of intersubband polaritons in dispersive cavities embedding GaAs/AlGaAs quantum wells are studied directly within their band structure using a non-collinear pump-probe geometry with phase-stable mid-infrared pulses. Selective excitation of the lower polariton at a frequency of ~25 THz and at a finite in-plane momentum, $k_\parallel$, leads to the emergence of a narrowband maximum in the probe reflectivity at $k_\parallel = 0$. A quantum mechanical model identifies the underlying microscopic process as stimulated coherent polariton-polariton scattering. These results mark an important milestone towards quantum control and bosonic lasing in custom-tailored polaritonic systems in the mid and far-infrared.**

Exciton polaritons [1], hybrid quasiparticles that emerge from the strong light-matter coupling between excitons and microcavity photons, have become a solid-state playground for fascinating phenomena previously reserved to ultracold atomic gases. Owing to their bosonic nature, exciton polaritons undergo stimulated scattering [2-5] and non-equilibrium Bose-Einstein condensation [6-11]. These processes have allowed the observation of exciting quantum phenomena [12], such as superfluidity [13] and hydrodynamic soliton nucleation [14], topological lasing [15-16], non-Hermitian effects [17] and spin-orbit interaction [18-20]. New materials such as transition metal dichalcogenides [21] and perovskites [22] have allowed for further customization of the potential landscape. However, in all these polaritonic systems the fundamental energy scale is intrinsically predetermined by the interband transition of the underlying material, limiting flexibility.

Conversely, the transition energy of intersubband (ISB) polaritons in semiconductor quantum well (QW) structures can be freely tuned by varying the QW width and doping [23] to reach the mid-IR and THz ranges of the electromagnetic spectrum [24-25]. Furthermore, the vacuum Rabi frequency, in this case, can be engineered to a substantial fraction of the transition energy [26-30], enabling novel quantum electrodynamical effects [31-36]. Ultrafast modulation of the coupling strength [28,37-39] opens intriguing possibilities for tailored sub-cycle dynamics [40]. Moreover, it has been predicted that ISB polaritons, like their excitonic counterpart, can undergo non-equilibrium condensation, enabling inversionless lasing, and thus on-chip coherent THz and mid-infrared light sources [41-43]. Yet up to now, a key ingredient given by a stimulated scattering mechanism towards the final state has been missing. First continuous-wave experiments have confirmed resonant ISB polariton-LO-phonon scattering towards the ground state [44]. However, the experiments revealed a spontaneous process, and not yet stimulation.

Here, we demonstrate ultrafast scattering between ISB polaritons under resonant excitation in a dispersive metal-insulator-metal resonator [45]. Our ultrafast non-collinear pump-probe setup allows for tunable narrowband pumping of the lower polariton (LP) branch at $k_{||} \neq 0$, while simultaneously probing both polariton branches at a different point of the band structure (normal incidence, $k_{||} = 0$). The experimental observations are reproduced very well by a quantum mechanical model of polariton-polariton scattering. This result constitutes a major step towards the realization of freely customizable quantum states of matter and polariton lasers in the mid-infrared and THz ranges.

The polaritonic sample under consideration consists of 36 repetitions of GaAs/Al$_{0.33}$Ga$_{0.67}$As semiconductor QWs with a well width of 8.3 nm and a nominal *n*-type doping concentration of $4.4 \times 10^{12}$ cm$^{-2}$ introduced as δ-layers in the center of the 20-nm-thick barriers. The QWs are embedded in a patterned metallic microcavity [44,45] with a top 1D grating featuring a lattice period of $\Lambda = 4.2$ μm [Fig. 1(a)]. By varying the angle of incidence of the incoming light, different in-plane wavevectors can be accessed. The ISB transition lies at an energy of $\hbar\omega_{\text{ISB}} = 0.12$ eV as confirmed by transmission measurements and calculations [Fig. 1(b)]. Figure 1(c) shows the numerically simulated polaritonic dispersion obtained with a rigorous coupled-wave analysis (RCWA). It fits the experimental data (dots) obtained from reflectivity measurements, as thoroughly described in Ref. [44]. Strong coupling between the dispersive cavity mode and the ISB transition leads to the emergence of two polariton branches observable as reflectance minima (blue regions). They exhibit a pronounced anticrossing at an in-plane wavevector of $k_{\parallel} = 0.37 \times \pi/\Lambda$ with a coupling strength of $2\Omega_R = 0.24\ \omega_{\text{ISB}}$. The LP has an inflection point at $k_{\parallel} = 0.29 \times \pi/\Lambda$.

Polariton interaction in exciton-polariton systems, in particular polariton-polariton scattering, has led to resonant amplification, when the system is pumped at the inflection point of the LP branch with a finite in-plane wavevector $k_{\parallel} = k_P$ [Fig. 1(e)] [2-5]. The simultaneous presence of a probe pulse leads to stimulated scattering, where two polaritons at the inflection point scatter under energy and momentum conservation to $k_{\parallel} = 0$ (signal) and $k_{\parallel} = 2 \times k_P$ (idler), respectively. Scattered signal polaritons can subsequently emit a photon in the direction of the probe pulse. For ISB polaritons, however, little is known about the dynamics on the picosecond timescale [37-39], and ISB polariton-polariton scattering transferring finite in-plane momenta has not been experimentally observed. To explore a scenario akin to what is reported for exciton polaritons, we excite the LP near the inflection point with a pump pulse while simultaneously monitoring the polariton population via a broadband probe pulse at a different wavevector and energy [Figs. 1(c) and 1(d)].

To this end, we developed a novel non-collinear mid-infrared pump-probe setup [Fig. 2(a)]. The sample is held in a helium-flow cryostat at a temperature of 10 K. Ultrabroadband, phase-locked probe pulses (duration, 68 fs), which contain frequency components between 15 THz and 45 THz above the noise floor, interrogate both LP and upper polariton (UP) under normal incidence, i.e., at $k_\parallel = 0$. The bandwidth-limited pump (FWHM, 3.9 THz) is centered at a frequency of 26.2 THz (Supplemental Material [46]). Goniometric beam steering allows the incidence angle to be tuned while maintaining temporal and spatial overlap with the probe pulse at the sample position. We choose a pump incidence angle of 21°, ($k_\parallel = 0.25 \times \pi/\Lambda$), which is close to the calculated inflection point of the LP. The diameter of the pump focus of 150 μm (intensity FWHM) ensures uniform excitation within the probe spot (diameter, 100 μm). The pump peak intensity at the sample surface reaches 350 MW/cm$^2$ (pump fluence, 43 μJ/cm$^2$), while the probe pulse is four orders of magnitude less intense. A germanium wafer placed inside the probe arm couples the probe reflected from the sample into a grating spectrometer. For quantitative reflectance spectra, we reference the probe spectrum with a signal reflected off a bare gold surface next to the polaritonic sample.

Figure 2(b) shows the reflectance of the sample at $k_\parallel = 0$ as a function of frequency and delay time $\Delta t$ between the pump and the probe pulse. For $\Delta t < 0$, i.e., before the arrival of the pump, the spectrum reveals two minima at frequencies of 24.2 THz and 30.4 THz, corresponding to the LP and UP states, respectively. Upon excitation by the pump pulse ($\Delta t = 0$), these resonances collapse within 400 fs and a new single minimum at 26.7 THz appears. The polariton splitting subsequently reemerges within a few picoseconds. To gain microscopic insights into the carrier dynamics, we model the experiment theoretically by expanding the framework of the optical Bloch equations to incorporate a dispersive microcavity, as employed in our experiment (Supplemental Material [46]). This approach, discussed in detail further below, allows us to extract the excited state population at the temporal center of the probe pulse in dependence of $\Delta t$ [Fig. 2(c)]. At its peak near $\Delta t = 0$, the excited population reaches a sizable fraction of 20% of all carriers. This is followed by an exponential decay of the population with a decay time of 2.5 ps. The significant number of excited carriers leads to the observed collapse of the coupling strength due to optical bleaching [37]. Indeed, the single resonance observed at $\Delta t = 0$ shows excellent agreement with the expected bare cavity resonance at 26.5 THz [Fig. 2(d)]. Importantly, the in-plane wavevectors of excitation and observation do not align in our experiment due to the non-collinear pump-probe geometry. Thus, the transient collapse of the coupling strength observed at $k_\parallel = 0$ provides evidence of the collapse of the entire polaritonic band structure, observed here for the first time.

The observation of transient bleaching and the extraction of the ISB carrier lifetime are important guardrails for our understanding of the ISB polariton dynamics. However, the full quench of the polariton bands is undesirable in the context of bosonic lasing, since ISB polaritons are composite particles which lose their bosonic properties under high excitation densities [41]. We therefore refine our experimental approach using lower peak intensities and a more selective excitation of the LP branch. To this end, the bandwidth of the pump pulses is reduced to an intensity FWHM of only 0.6 THz (Supplemental Material [46]), even smaller than the polariton linewidth [Fig. 1(d)].

Figure 3(a) shows the time-averaged reflectance of the sample at normal incidence, $k_\parallel = 0$, as a function of probe frequency and delay time $\Delta t$. The pump fluence at the sample surface of 39 µJ/cm$^2$, corresponding to a peak intensity of 70 MW/cm$^2$, is still enough to induce a shift of the UP and a simultaneous decrease in reflectance in the spectral region between the polariton branches. However, no reflectance minimum at the spectral position of the cavity mode appears. This attests to an incomplete bleaching of the ISB polariton system as opposed to the full collapse observed in Fig. 2. Most remarkably and in contrast to the observations in Fig. 2, a local reflectance maximum emerges at the high-frequency side of the LP around $\Delta t = 0$. This new peak features a center frequency of 24.8 THz and a spectral and temporal width of approximately 0.6 THz and 1 ps, respectively. The limited duration signifies that the emergence of this local maximum requires the presence of both pump and probe pulses. We will show in the following that this signature is the fingerprint of coherent scattering of polaritons at differing in-plane momenta. To this end, we describe the dispersive microcavity and its wavevector-dependent nonlinear interaction in terms of an extended optical Bloch equations framework involving several momentum states.

The theoretical description includes the dynamics of the electromagnetic field within the dispersive microcavity and of a spatially distributed ISB transition within the QW layer (Supplemental Material [46]). Given the (discrete) translational invariance along the cavity plane, photonic modes form a Bloch band with a frequency dispersion $\omega_k$ as a function of the in-plane quasi-wavevector $k_\parallel$. Nonlinear effects can couple polariton states at differing wavevectors, calling for a multi-mode description. The simplest model of the pump-probe experiment includes three cavities with frequencies $\omega_P^{cav}, \omega_S^{cav}$, and $\omega_I^{cav}$ and corresponding wavevectors $k_{P,S,I}$. Here, $k_P$ and $k_S$ denote the incident wavevector of the pump and probe pulses, respectively, while the additional cavity at ($\omega_I^{cav}$, $k_I = 2k_P - k_S$) ensures energy and momentum conservation as the polaritons at $k_P$ and $k_S$ interact. Within a semiclassical picture, the intracavity field $E(x)$ at position $x$ takes the form $E(x) = E_P e^{ik_P x} + E_S e^{ik_S x} + E_I e^{ik_I x}$.

In an analogous picture, the ISB transition is described in terms of a spatial distribution of two-level emitters with spatially dependent polarization $\sigma(x)$ and population difference $\Pi(x)$ between the two electronic subbands. These quantities can be expanded as $\sigma(x) = \sigma_P e^{ik_P x} + \sigma_S e^{ik_S x} + \sigma_I e^{ik_I x}$ and $\Pi(x) = \Pi_0 + \Pi_q e^{iqx} + \Pi_{-q} e^{-iqx}$, where $q = k_P - k_S$. The polarization $\sigma(x)$ contains components corresponding to the three momentum states of the intracavity field, while the population difference $\Pi(x)$ consists of a uniform population $\Pi_0$ plus spatial modulations $\Pi_{q,-q}$ owing to the interference of the different intracavity field components. The time evolution of our system is then described by nine ordinary differential equations for the three components of $E$, $\sigma$ and $\Pi$. The intracavity fields evolve according to

$$i\partial_t E_{P/S/I} = \left(\omega_{P/S/I}^{cav} - \frac{i}{2}\Gamma_{cav}\right) E_{P/S/I} - \Omega_R \sigma_{P/S/I} + E_{P/S/I}^{ext}(t),$$

with the cavity damping rate $\Gamma_{cav}$. The externally incident fields $E_{P/S}^{ext}$ are the waveforms of the pump and probe fields, respectively, and $E_I^{ext} = 0$. The intrinsic nonlinearity of the optical Bloch equations leads to coupling between different momentum states, e.g.,

$$i\partial_t \sigma_S = \omega_{ISB}\,\sigma_S + \Omega_R(E_S\Pi_0 + E_P\Pi_{-q}) + \eta\big[(\Pi_0 + 1)\sigma_S + \Pi_{-q}\sigma_P\big] - i\frac{\Gamma_{coh} + \Gamma_{EID}(\Pi_0 + 1)}{2}\sigma_S.$$

The mode mixing originates from light scattering on the spatially-modulated population difference $\Pi(x)$ via the nonlinear saturation of the ISB transition and the nonlinear suppression of the depolarization shift included as $\eta$ [53]. The coupled differential equations are numerically solved for a set of parameters of $\omega_{cav}, \omega_{ISB}, \Omega_R, \Gamma_{coh} = 18$ THz and $\Gamma_{cav} = 10$ THz retrieved from the steady-state characterization of the sample, while excitation-induced dephasing $\Gamma_{EID} = 20$ THz and $\eta = -2\pi \times 4.8$ THz are chosen for optimum agreement with the experiment (Supplemental Material [46]).

The calculated reflectance shown in Fig. 3(b) reproduces both the saturation and the local maximum near the LP very well. Comparing spectra for $\Delta t = 0$ [Fig. 3(c)] even reveals quantitative agreement between experiment (blue line) and theory (shaded area) regarding center frequency, spectral shape, and absolute reflectance. The spectral width of the feature is comparable to the width of the pump pulses (red line). Notably, the local maximum is red-shifted with respect to the pump frequency in both experiment and theory.

Our theoretical model allows us to further determine the microscopic origin of the local reflectance maximum. Nonlinearities between the three different polariton states are mediated by the spatially modulated components $\Pi_q$ and $\Pi_{-q}$ of the population difference. This can be unequivocally tested by holding $\Pi_q = \Pi_{-q} = 0$ while keeping all other parameters as in Fig. 3(b). In the resulting spectra [Fig. 3(d)], the local maximum vanishes completely, manifesting that polariton-polariton interactions are at the origin of the signature. To the best of our knowledge these experiments mark the first observation of ultrafast scattering between ISB polaritons with different wavevectors.

The observed novel process could be similar to the known non-dissipative, i.e. parametric, scattering of exciton polaritons sketched in Fig. 1(e): two pump polaritons at $k_\| = k_P$ can scatter under energy and momentum conservation to $k_\| = 0$ and $k_\| = 2 \times k_P$, respectively, when stimulated by signal polaritons created by the probe pulse. Polaritons scattered to $k_\| = 0$ can then emit light into the probe pulse at $\omega_S^{cav}$.

Additionally, in a broadened band structure, dissipative scattering processes of polaritons off the population grating formed by the pump and probe pulses can contribute to the scattered signal (Supplemental Material [46]). The relative strength of the two microscopic processes depends on the specific configuration, e.g. on the polariton dispersion and the linewidths.

We further characterize the polariton-polariton scattering quantitatively by investigating its pump power dependence. Figures 4(a)-(c) show reflectance spectra at the LP for different pump peak intensities $I_\text{peak}$ of $40\,\text{MW/cm}^2$ (a), $57\,\text{MW/cm}^2$ (b), and $80\,\text{MW/cm}^2$ (c). The scattering signal maintains its spectral and temporal position, reaching a reflectance as high as 0.5 for the highest pump fluence [Fig. 4(d)]. Additionally, the local maximum is monitored when varying the pump peak intensity continuously between a few kW/cm² and $40\,\text{MW/cm}^2$ [Fig. 4(e), black line]. For $I_\text{peak} < 4\,\text{MW/cm}^2$, the efficiency of the polariton-polariton scattering process scales superlinearly with respect to the pump intensity. For higher pump intensities, saturation is observed, most likely in connection to the ISB transition bleaching as discussed above. The quantum mechanical theory reproduces both the superlinear increase of the reflectance and, within a factor of 2.5, the subsequent saturation for high pump powers [Fig. 4(g)]. Conversely, the signal intensity scales linearly as a function of the probe pulse intensity [Fig. 4(f)], as expected for a scattering process involving bosonic particles, thus signifying the stimulated character of the process. Finally, we note that only a fraction of the scattered polariton population decays radiatively out of the cavity. While, as of now, saturation occurs below a reflectance value of 1, further improvements, for example in the quality factor of the ISB transition [54], could well enable gain in the near future.

In conclusion, we have investigated the ultrafast dynamics of ISB polaritons in a dispersive metal-insulator-metal cavity in a non-collinear mid-infrared pump-probe experiment. At large fluences, phase-stable ultrashort pulses bleach the ISB transition on a timescale of a few 100 fs, inducing a full collapse of the coupling strength. Conversely, when the LP is selectively excited with narrowband pump pulses, ISB polaritons undergo stimulated scattering between states at finite in-plane momentum and the bottom of the LP band at $k_\parallel = 0$. In an intermediate excitation regime below saturation, the efficiency of this process increases nonlinearly with the pump power. We believe that with further optimization – for example, by fine-tuning the Hopfield coefficients of the polaritonic states involved (Supplemental Material [46]) and by designing resonators with lower losses [55] – this process could enable gain and eventually condensation and inversionless lasing of ISB polaritons [37-39].

We thank Martin Furthmeier, Ignaz Laepple and Imke Gronwald for technical assistance, and S. De Liberato for discussions. This work was supported by the European Union Future and Emerging Technologies (FET) Grant No. 737017 (MIR-BOSE). This work was partially supported by the French RENATECH network.

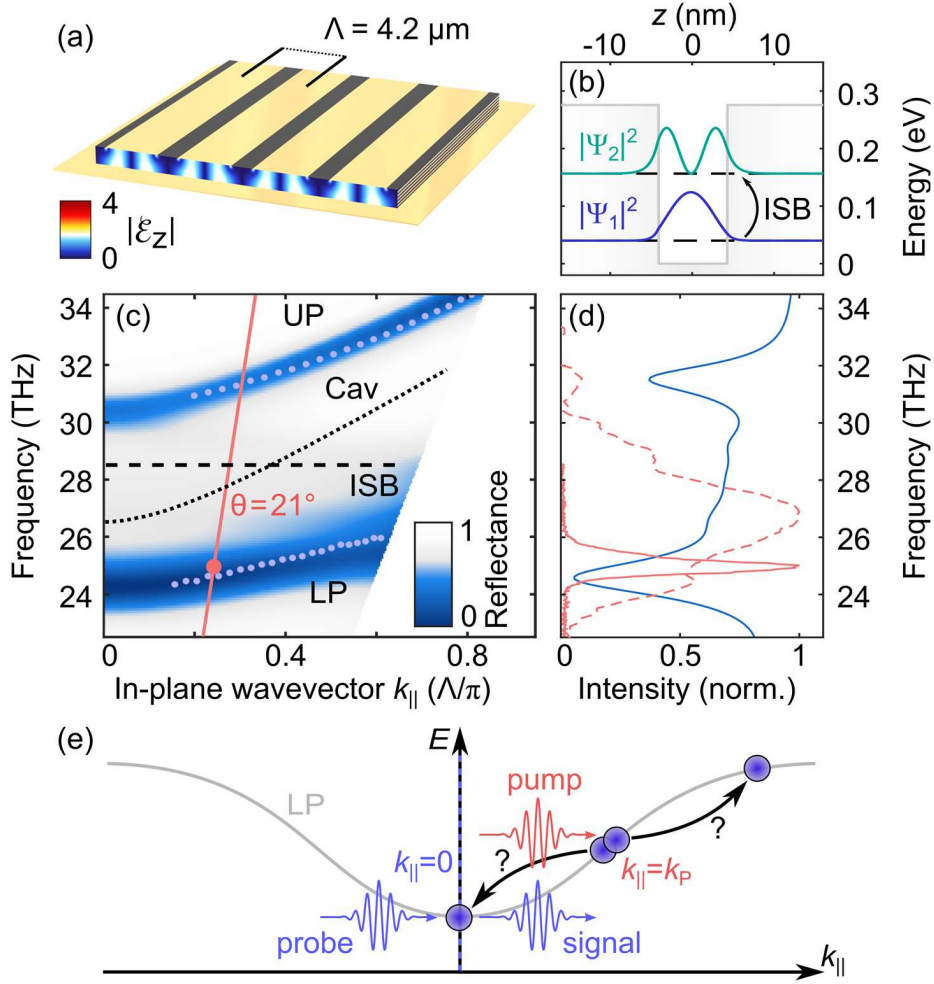

Figure 1 (a) Sample scheme and spatial distribution of the out-of plane electric field component $|\mathcal{E}_z|$. Λ: metal grating periodicity. (b) Conduction band structure (grey line) and wavefunction squared moduli for first ($\Psi_1$) and second ($\Psi_2$) subbands inside an 8.3 nm GaAs/Al$_{0.33}$Ga$_{0.67}$As QW with growth direction $z$. ISB: intersubband transition. (c) Calculated reflectance (RCWA) of the sample. The strong coupling of ISB (black dashed) and cavity resonance (black dotted) leads to the emergence of polariton states (LP and UP). Measured data (blue dots) is shown for comparison. The red line marks the pump angle of incidence θ = 21°. (d) Spectra of narrowband (red solid line) and broadband (red dashed line) pump pulses and calculated reflectance at θ = 21° (blue line) at a lattice temperature of 10 K. (e) Sketch of parametric polariton-polariton scattering: The pump pulse creates polaritons at finite $k_{||}$. Scattering of two pump polaritons leads to amplification of the probe pulse through one of the polaritons, while excess momentum is carried away by the other polariton.

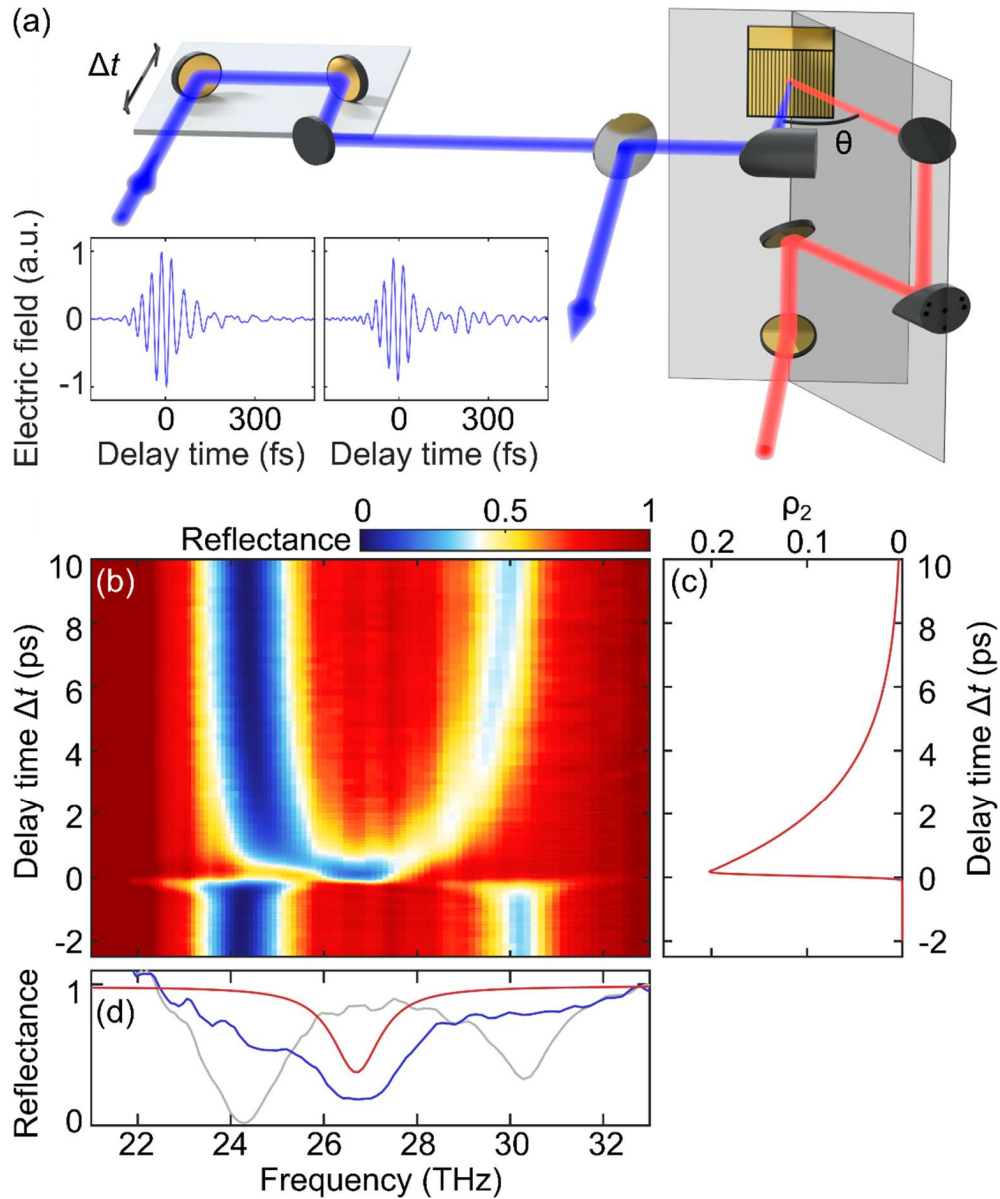

Figure 2 (a) Schematic geometry of the angle-resolved ultrafast pump-probe experiment. The sample is probed at normal incidence with varying pump-probe delay time $\Delta t$. The angle of incidence θ of the pump pulses can be adjusted without changing $\Delta t$ or the spatial overlap of the pump and probe foci. Inset: Measured waveforms as incident onto the sample (left) and reflected from it (right). (b) Measured reflectance spectra as a function of pump-probe delay time, $\Delta t$, for a broadband pump pulse with a bandwidth of 3.9 THz (peak intensity 350 MW/cm$^2$, θ = 21°). (c) Theoretical calculation of the excited polariton population $\rho_2$. (d) Spectra before the arrival of the pump pulse (grey) and at $\Delta t = 0$ (blue), and calculated reflectance of the uncoupled cavity resonance (red).

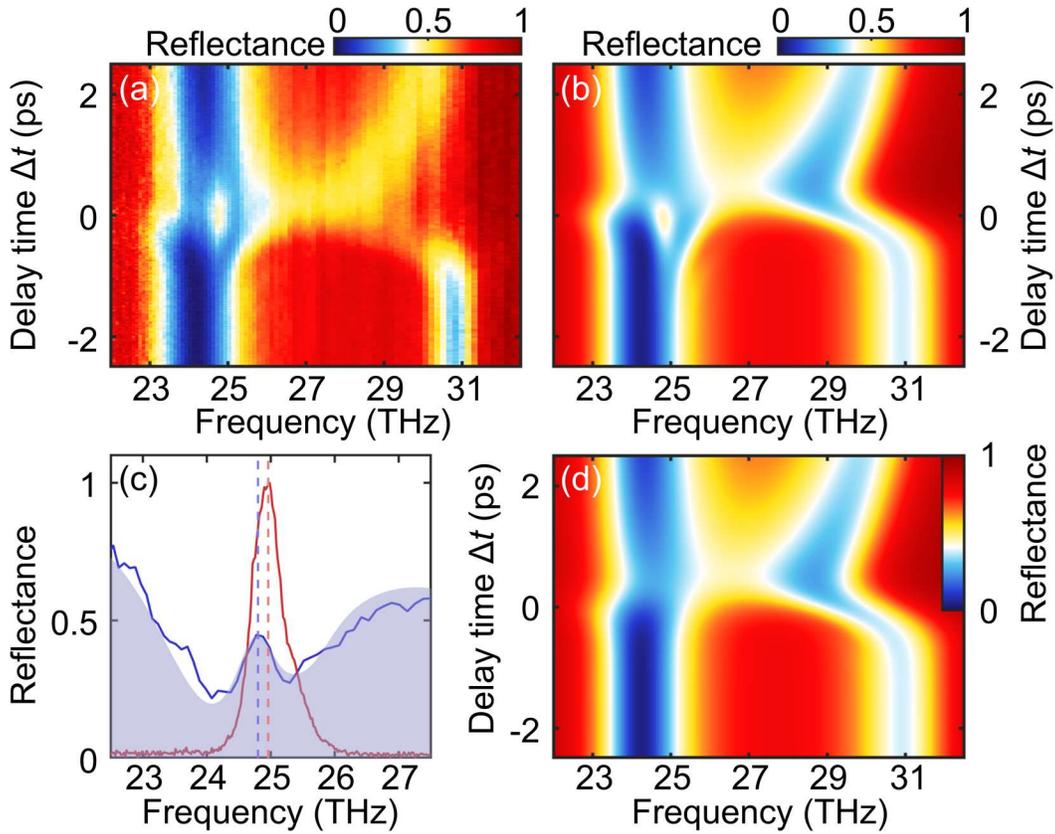

Figure 3 (a) Measured reflectance spectra in dependence of pump-probe delay time, $\Delta t$, for a narrowband pump pulse with a bandwidth of 0.6 THz (peak intensity 70 MW/cm$^2$, $\theta = 21°$). (b) Simulated reflectance including scattering terms between pump and probe pulses. (c) Experimental pump spectrum (red) as well as measured (blue line) and simulated (shaded area) reflectance at $\Delta t = 0$. A narrowband feature emerges with slightly lower center frequency (blue dashed line, 24.8 THz) than the pump center frequency (red dashed line, 25 THz). (d) Simulated reflectance excluding scattering terms between pump and probe pulses.

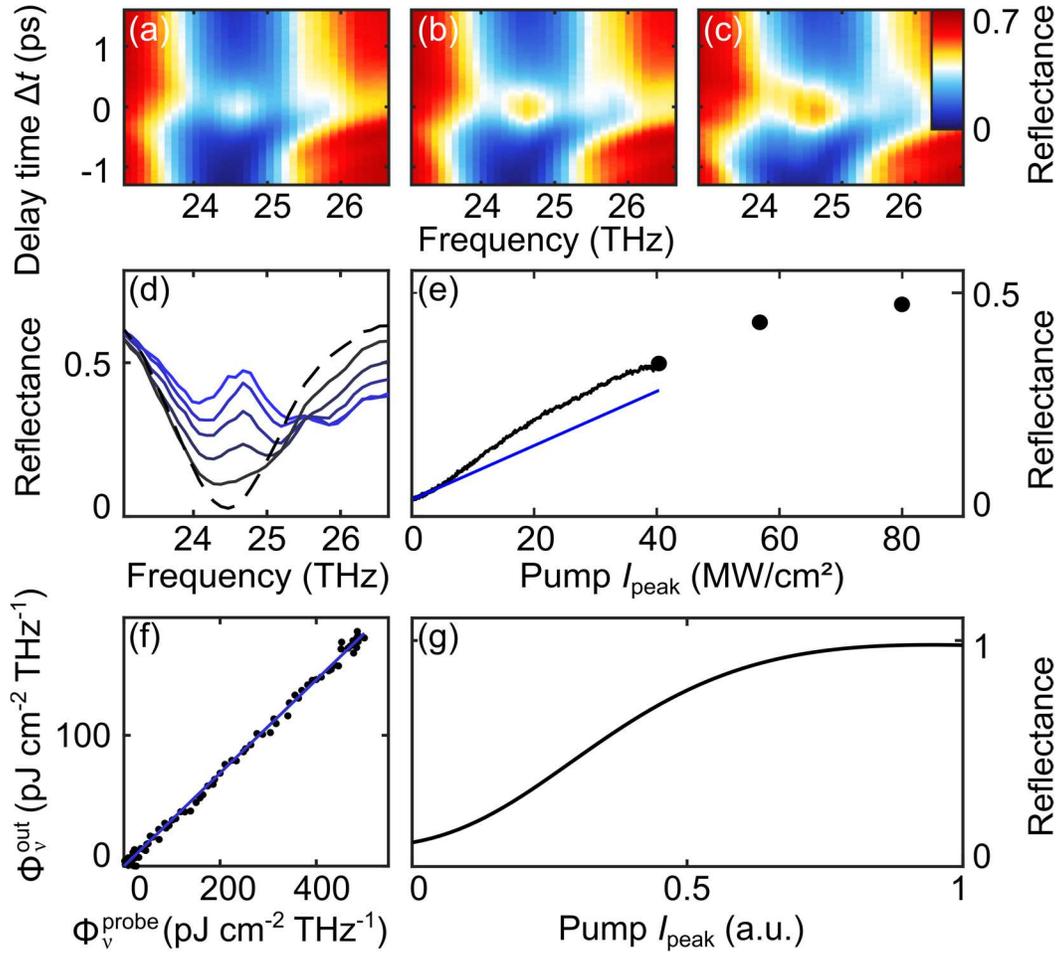

Figure 4 (a-c) Measured reflectance spectra for pump peak intensities of 40 MW/cm$^2$ (a), 57 MW/cm$^2$ (b), and 80 MW/cm$^2$ (c) as a function of $\Delta t$. (d) Reflectance for different pump intensities ranging from 10 MW/cm$^2$ to 80 MW/cm$^2$ at $\Delta t = 0$ (solid lines) and for $\Delta t < 0$ (dashed). (e) Black curve: Reflectance at the spectral maximum of the signature at $\Delta t = 0$ as a function of pump peak intensity $I_{peak}$. Black dots: Discrete values extracted from measurements shown in figure parts (a-c). Blue line: Linear fit to the data for pump intensities between 180 kW/cm$^2$ and 4 MW/cm$^2$. (f) Measured spectral fluence $\Phi_\nu^{out}$ in dependence of probe input $\Phi_\nu^{probe}$ at the spectral maximum of the signature at $\Delta t = 0$ (black dots) and linear fit (blue line). The linear dependence signifies stimulation. (g) Simulated reflectance for $\Delta t = 0$ at the spectral maximum of polariton-polariton scattering in dependence of the pump intensity.

19